\documentclass{PoS}

\title{Renormalization of the energy-momentum tensor on the lattice}

\ShortTitle{Renormalization of the energy-momentum tensor on the lattice}

\author{\speaker{Michele Pepe}\\
        INFN, Sezione di Milano-Bicocca\\ 
        Edificio U2, Piazza della Scienza 3\\ 
        20126 Milano, Italy.\\
        E-mail: \email{Michele.Pepe@mib.infn.it}}
\author{Leonardo Giusti\\
        Dipartimento di Fisica, Universit\`a di Milano-Bicocca\\
        and INFN, sezione di Milano-Bicocca\\
        Edificio U2, Piazza della Scienza 3\\ 
        20126 Milano, Italy.\\
       E-mail: \email{Leonardo.Giusti@mib.infn.it}}

\abstract{We present the calculation of the non-perturbative renormalization constants of the
  energy-momentum tensor in the $SU(3)$ Yang-Mills theory. That computation is carried out
  in the framework of shifted boundary conditions, where a thermal quantum field theory
  is formulated in a moving reference frame. The non-perturbative renormalization factors
  are then used to measure the Equation of State of the $SU(3)$ Yang-Mills theory. Preliminary
  numerical results are presented and discussed.}

\FullConference{The European Physical Society Conference on High Energy Physics\\
                 22-29 July 2015\\
                 Vienna, Austria}

\begin{document}
\section{Introduction}

The energy-momentum tensor is a fundamental field of a Quantum Field Theory since it
contains the currents associated to Poincar\'e symmetry and scale invariance. Furthermore,
the expectation values of its matrix elements are physical quantities since they are directly related to the thermodynamic features of
a quantum theory at finite temperature~\cite{Landau1987}. When the regularization of a quantum theory
preserves space-time symmetries --~like, for instance, dimensional regularization~-- the
energy-momentum tensor does not renormalize since it is related to a conserved current.
However, non-perturbative investigations starting from first principles are usually
carried out by considering the formulation of the quantum theory on the lattice.
This regularization explicitly breaks the Poincar\'e invariance that is recovered
only in the continuum limit. Thus, the energy-momentum tensor on the lattice is no longer
trivially a conserved current and it needs to be properly renormalized to guarantee
that the associated currents generate translations and rotations in the
continuum limit~\cite{Caracciolo:1989pt}. For the Yang-Mills theory, scale invariance is also broken by the
regularization; however, that symmetry is anomalous and it is not restored in the continuum
limit, generating a dynamical mass-gap.

The approach that has been used to define a properly renormalized energy-momentum tensor on
the lattice is to impose the validity of some Ward Identities at fixed lattice spacing up
to terms that vanish when the lattice spacing goes to 0~\cite{Caracciolo:1989pt}. Based on that framework, the
renormalization of the energy-momentum tensor has been computed in perturbation theory at
1 loop~\cite{Caracciolo:1991cp}. Although one can in principle construct a lattice
definition of the energy-momentum tensor, the non-perturbative calculation of the
renormalization factors can be not straightforward if one has to consider correlation
functions that are difficult to measure by numerical simulations. 

A few years ago -- based on an old suggestion by Landau~\cite{Landau1987} -- it has been proposed to
consider a thermal quantum field theory in a moving reference
frame~\cite{Giusti:2010bb,Giusti:2011kt,Giusti:2012yj}. In the path-integral language,
this can be accomplished by considering a spatial shift ${\bf \xi}$ when closing the
boundary conditions along the temporal direction. The shift ${\bf \xi}$ corresponds to the
Wick rotation of the speed of the moving frame. Interestingly, in this new framework, the
parity symmetry is explicitly broken by the shift. This allows to write down new Ward
Identities involving the energy-momentum tensor and new equations to measure both
thermodynamic quantities and the renormalization factors. In particular, an off-diagonal
matrix element of the energy-momentum tensor may have a non vanishing expectation value.
Numerical simulations with shifted boundary conditions have already provided new successful,
simple methods to study the thermodynamics of the Yang-Mills theory~\cite{Giusti:2014ila}. 

This report is organized as follows. In section 2, the main equations with shifted
boundary conditions are summarized both in the continuum and on the lattice. The next
section presents the results of the non-perturbative calculation of the renormalization
factors of the energy-momentum tensor and, in section 4, the renormalization of the
energy-momentum tensor is used to compute the Equation of State of the $SU(3)$ Yang-Mills
theory. Conclusions and outlook follow.

\section{Shifted boundary conditions and Ward Identities}

In this section we consider the $SU(3)$ Yang-Mills theory at finite temperature in the
Euclidean space. The path-integral formulation in a moving reference frame can be
defined by considering  shifted boundary conditions~\cite{Giusti:2012yj}
\begin{equation}\label{eq:shfbc}
A_\mu(L_0,\vec x) =A_\mu(0,\vec x - L_0\vec\xi)  
\end{equation}
along the temporal direction of length $L_0$ with shift $\vec \xi \in\mathbb{R}^3$.
When $\vec \xi \neq 0$, the parity symmetry is broken and there are new interesting Ward Identities
involving the energy-momentum tensor $T_{\mu\nu}$~\cite{Giusti:2010bb,Giusti:2011kt,Giusti:2012yj} ($x_0\neq 0$):
\begin{equation}\label{eq:dxi}
L_0 \langle \; T_{0k} \rangle_{\vec\xi} = \frac{1}{V}
\frac{\partial}{\partial \xi_k} \ln Z(L_0,\vec \xi)\;, \qquad
L_0 \langle \; {\overline T}_{0k}(x_0) \, O(0) \rangle_{\vec\xi,\, c}
= \frac{\partial}{\partial \xi_k} \langle O \rangle_{\vec\xi}\;,  
\end{equation}
where $Z(L_0,\vec \xi)$ is the partition function with shifted boundary conditions and $O$
is a generic gauge invariant operator. The subscript $c$ 
indicates a connected correlation function, $\langle \cdot \rangle _{{\vec\xi}}$ stands
for the expectation value with shifted boundary conditions and ${\overline T}_{\mu\nu}=\int d^3x\,
T_{\mu\nu}(x)$. The field $T_{\mu\nu}$ can be defined by
\begin{equation}
T_{\mu\nu} (x) = \frac{1}{g_0^2} 
\left[ F_{\mu\rho}^a (x) F_{\nu\rho}^a (x)  
-\frac{1}{4} \delta_{\mu\nu} F_{\rho\sigma}^a (x) F_{\rho\sigma}^a (x) \right]
\end{equation}

\noindent
where $g_0$ is the bare coupling constant and the field strength is given in terms of the
gauge field $A_\mu (x)$ by 
$F_{\mu\nu}(x) = \partial_\mu A_\nu (x) - \partial_\nu A_\mu (x) -i [A_\mu (x),A_\nu (x)]$.
Other useful equations are the following~\cite{Giusti:2012yj,Giusti:2015daa}
\begin{equation}\label{eq:WIodd}
\langle T_{0k} \rangle_{\vec\xi} = \frac{\xi_k}{1-\xi_k^2} 
\left\{\langle T_{00} \rangle_{\vec\xi}  
- \langle T_{kk} \rangle_{\vec \xi}\,\right\}\; ,
\qquad
\frac{\partial}{\partial \xi_k} \langle T_{\mu\mu} \rangle_{\vec\xi} =
\frac{1}{(1+\xi^2)^2}\frac{\partial}{\partial \xi_k}
\left[\frac{(1+\xi^2)^3}{\xi_k} \langle T_{0k} \rangle_{\vec\xi} \right]\; . 
\end{equation}

When we consider the lattice regularization, the 10 dimensional symmetric $SO(4)$ representation
of the energy-momentum tensor  becomes reducible and it splits into the sum
of the singlet, the triplet, and the sextet representations of the hyper-cubic group. The field
$T_{\mu\nu}$ can then be expressed as a combination of the following three operators
\begin{equation}\label{eq:lattmunu}
T^{[1]}_{\mu\nu} = (1-\delta_{\mu\nu}) \frac{1}{g_0^2} F^a_{\mu\alpha}F^a_{\nu\alpha}; \quad
T^{[2]}_{\mu\nu} = \delta_{\mu\nu}\, \frac{1}{4 g_0^2}\, F^a_{\alpha\beta} F^a_{\alpha\beta} ;\quad
T^{[3]}_{\mu\nu} = \delta_{\mu\nu} \frac{1}{g_0^2} \Big\{F^a_{\mu\alpha}F^a_{\mu\alpha} 
- \frac{1}{4} F^a_{\alpha\beta}F^a_{\alpha\beta} \Big\}
\end{equation}

\noindent
and the identity. Since translation and rotation symmetries are broken by the lattice
regularization, the sextet $T^{[1]}_{\mu\nu}$ and the triplet $T^{[3]}_{\mu\nu}$ operators
pick up a multiplicative renormalization factor, while the singlet $T^{[2]}_{\mu\nu}$
mixes in addition with the identity.
The renormalized energy-momentum tensor can finally be written as
$ T^{\rm R}_{\mu\nu} = Z_{_T} \Big\{T^{[1]}_{\mu\nu} + z_{_T}  T^{[3]}_{\mu\nu} + 
z_{_S} \big[T^{[2]}_{\mu\nu} - 
\langle T^{[2]}_{\mu\nu} \rangle_0 \big]\Big\}$ .

Because of the finite renormalization factors, we can write the lattice version of the first
equation of (\ref{eq:dxi}) as follows
\begin{equation}\label{ZTpractic}
Z_{_T}(g_0^2) = - \frac{\Delta f}{\Delta \xi_k}\, \frac{1}{\langle T^{[1]}_{0k} \rangle_{\vec\xi} }\; , 
\qquad \mbox{with} \quad
\frac{\Delta f}{\Delta \xi_k} = \frac{1}{2a V}
\ln\Big[\frac{Z(L_0,{\vec\xi }-a \hat k/L_0 )}{Z(L_0,{\vec\xi}+a \hat k/L_0 )} \Big] .
\end{equation}
The two equations in (\ref{eq:WIodd}) are given by
\begin{equation}\label{Zd}
z_{_T}(g_0^2) = \frac{1-\xi_k^2}{\xi_k} \frac {\langle T^{[1]}_{0k}
  \rangle_{\vec\xi} }{\langle T^{[3]}_{00}\rangle_{\vec\xi} - \langle T^{[3]}_{kk} \rangle_{\vec\xi}}\; ,
\end{equation}
\begin{equation}\displaystyle
z_{_S} = \frac{1}{(1+\xi^2)^2}\frac{\left[\frac{(1+\xi^{'2})^3}{\xi'_k} \langle T^{[1]}_{0k} \rangle_{\vec\xi'} 
\right]_{\vec\xi'=\vec\xi+ a \hat k/L_0} - 
\left[\frac{(1+\xi^{'2})^3}{\xi'_k} \langle T^{[1]}_{0k} \rangle_{\vec\xi'} 
\right]_{\vec\xi'=\vec\xi- a \hat k/L_0}
}
{\langle T^{[2]}_{\mu\mu} \rangle_{\vec\xi+ a \hat k/L_0} - \langle T^{[2]}_{\mu\mu} \rangle_{\vec\xi- a \hat k/L_0}}\; . 
\end{equation}

Note that the equations above allow for a fully non-perturbative definition of
$T_{\mu\nu}$; furthermore they represent a simple procedure to perform the numerical
calculation of $Z_{_T}$, $z_{_T}$ and $z_{_S}$ since only the expectation values of local
fields need to be measured. 

\section{Numerical results}

In this section we present the results of Monte Carlo simulations to calculate $Z_{_T}
(g_0^2)$ and $z_{_T} (g_0^2)$ in the range $g_0^2\in(0,1)$~\cite{Giusti:2015daa}; work is in progress
for the computation of  $z_{_S} (g_0^2)$. We define the SU(3) Yang--Mills theory on a
space-time lattice of volume $L^3\times L_0$ and lattice spacing $a$. We impose periodic boundary conditions in the
spatial directions and shifted boundary conditions in the temporal direction:
$U_\mu(L_0,\vec x) =U_\mu(0,\vec x - L_0\vec\xi)$, where $U_\mu(x_0,\vec x)$ are the link variables. 
We consider the standard Wilson action $S[U] = -1/g_0^2\, \sum_{x,\mu\nu} {\rm Re}\mbox{Tr} [ U_{\mu\nu}(x) ]$,
where the plaquette is given by
$U_{\mu\nu}(x) = U_\mu(x)\, U_\nu(x+ a\hat \mu)\, U^\dagger_\mu(x + a\hat \nu)\,
U^\dagger_\nu(x)$. The gluon field strength tensor is defined as~\cite{Caracciolo:1989pt}  
\begin{equation}
F^a_{\mu\nu}(x) = - \frac{i}{4 a^2} 
\mbox{Tr} \Big\{\Big[Q_{\mu\nu}(x) - Q_{\nu\mu}(x)\Big]T^a\Big\}\; , 
\end{equation}
where $Q_{\mu\nu}(x) = U_{\mu\nu}(x) + U_{\nu-\mu}(x) + U_{-\mu-\nu}(x) + U_{-\nu\mu}(x)$,
and the minus sign stands for the negative orientation. The renormalization constants $Z_{_T}$, $z_{_T}$ and $z_{_S}$ are finite 
and depend on $g^2_0$ only. Considering the above definition of the field strength tensor
on the lattice, at 1 loop in perturbation theory their expressions are~\cite{Caracciolo:1989pt,Caracciolo:1991cp}
\begin{equation}\label{eq:PT1loop}
Z_{_T}(g_0^2) = 1 + 0.27076 \; g_0^2\;, \quad
z_{_T}(g_0^2) = 1 - 0.03008\; g_0^2 \;, \quad 
z_{_S}(g_0^2) = \frac{b_0}{2} g_0^2\; . 
\end{equation}

\subsection{Computation of $Z_{_T}$\label{sec:Ztmeth}}
\begin{figure}[htb]
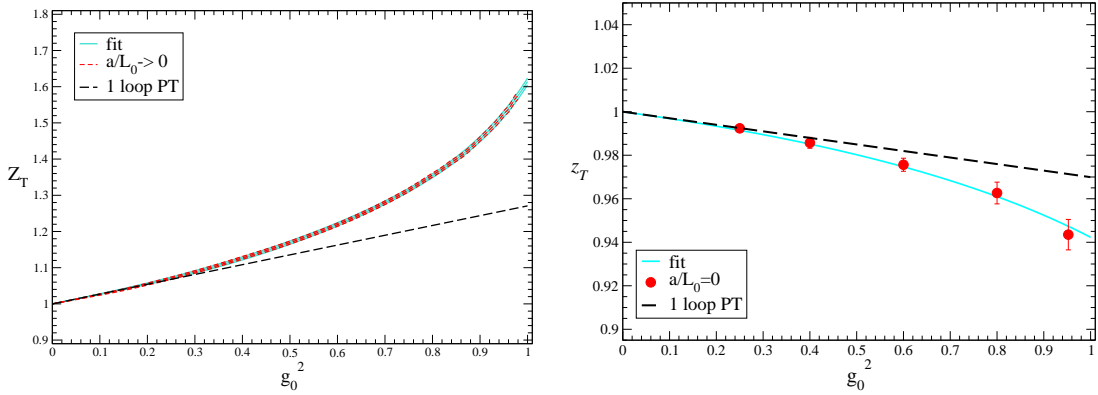

\includegraphics[width=7.0 cm,angle=0]{ZTfit.eps}\ \ \ \ \ 
\includegraphics[width=7.0 cm,angle=0]{zts.eps}
\caption{
The renormalization factor $Z_{_T}(g_0^2)$ (left panel) and $z_{_T}(g_0^2)$ (right panel)
as a function of the bare coupling $g_0^2$. The dashed lines represent the 1-loop
perturbative results and the solid lines are interpolating fits of the numerical data.\label{fig:ZT}}
\end{figure}

The direct determination of $\Delta f/\Delta \xi_k$ in (\ref{ZTpractic}) is a numerically
challenging problem since it requires the computation of the ratio of two partition
functions with a poor overlap of the relevant phase space~\cite{deForcrand:2000fi,DellaMorte:2007zz,Giusti:2010bb}.  
Moreover, the calculation becomes quickly demanding for large lattices because the numerical
cost increases quadratically with the spatial volume. Since $\Delta f/\Delta \xi_k$ is a
smooth function of $g_0^2$ at fixed values of $L_0/a$ and $L/a$ in the range of chosen
values, its derivative with respect to $g_0^2$ can be written as  
\begin{equation}\label{intder}
\frac{d}{d g_0^{2}} \frac{\Delta f}{\Delta \xi_k} 
= \frac{1}{2a L^3 g_0^2}\,
\Big\{\langle S \rangle _{{\vec\xi} - a/L_0 \hat k} -\langle S \rangle _{{\vec\xi} + a/L_0 \hat k}\Big\}\; .
\end{equation}
The difference in the r.h.s.
has been computed for ${\vec\xi}=(1,0,0)$ and $L/a=48$ at $L_0/a=3,4$ and $5$ for many
values of $g_0^2$. At each value of $L_0/a$ the points are 
interpolated with a cubic spline, and the resulting curve is integrated over $g_0^2$. The 
free-case value is computed analytically and is added to the integral. Then $\langle
T^{[1]}_{0k} \rangle_{\vec\xi}$ has also been computed at many
values of $g_0^2$ and the results have been interpolated with cubic splines. The final
result for $Z_{_T}$ is shown in the left panel of Fig.~\ref{fig:ZT} together with the
1-loop perturbative result and an interpolating fit
\begin{equation}\label{eq:ZTfinal}
Z_{_T}(g_0^2) = \frac{1 - 0.4457\, g_0^2}{1 - 0.7165\, g_0^2} - 0.2543\, g_0^4 
               + 0.4357\, g_0^6 - 0.5221\, g_0^8
\end{equation}

\subsection{Determination of $z_{_T}$}
The renormalization constant $z_{_T}$ is calculated by 
imposing the tree-level improved version of Eq.~(\ref{Zd}) given by
$\{ z_{_T}(g_0^2) - \mbox{ free case}\}$ , with
$\frac{L\, \xi_k}{L_0(1+\xi_k^2)} = q \in \mathbb{Z}$. The expectation values of  
$\langle T^{[1]}_{0k} \rangle_{\vec\xi}$ and of the difference 
$\langle T^{[3]}_{00} \rangle_{\vec\xi} - \langle T^{[3]}_{kk} \rangle_{\vec\xi}$ are 
measured straightforwardly in the same simulation.

We chose ${\vec\xi} = (1/2,0,0)$ and $q=8$ so that 
the ratio of the spatial linear size over the temporal 
one is fixed to be $L/L_0=20$. We simulated 5 values of $g_0^2$ in the range 
$0\leq g_0^2 \leq 1$ with temporal length $L_0/a=4,6,8$ and $12$. After performing a
combined extrapolation to $a/L_0=0$, the final results are shown in the right panel of
Fig.~\ref{fig:ZT}. The dashed line is the 1-loop perturbative result and the solid
line is an interpolating fit
\begin{equation}\label{eq:ztsfinal}
z_{_T}(g_0^2) = \frac{1 - 0.5090\, g_0^2}{1 - 0.4789\, g_0^2}\; , 
\end{equation}

\section{A physical application: the Equation of State}
In this section we use the results of the non-perturbative renormalization of the
energy-momentum tensor to obtain the Equation of State from Monte Carlo simulations. The
energy-momentum tensor is a physical field and its expectation values are related to the thermodynamics
features of a quantum field theory. In fact an interesting Ward Identity that follows using shifted boundary
conditions is the following~\cite{Giusti:2012yj}

\begin{equation}\label{EoS}
\frac{s}{T^3}=  -
\frac{L_0^4 (1+{\bf \xi} ^2)^3}{ \xi_k}  \langle T_{0k} \rangle_\xi Z_T,
\end{equation}
where $s$ is the entropy density and $T=1/L_0\sqrt{1+{\bf \xi} ^2}$ is the temperature. In
\cite{Giusti:2014ila} the temperature dependence of $s/T^3$ has been measured using the step-scaling
function: in that approach one can avoid computing $Z_T$ but only fixed, constant steps in
the temperature can be done. Once the renormalization factor $Z_T$ is known, the
eq.~(\ref{EoS}) allows to measure $s/T^3$ independently at any temperature; furthermore the
extrapolation to the continuum limit is carried out in a very simple way. 

\begin{figure}[htb]
\centering
\includegraphics[width=12.0 cm,angle=0]{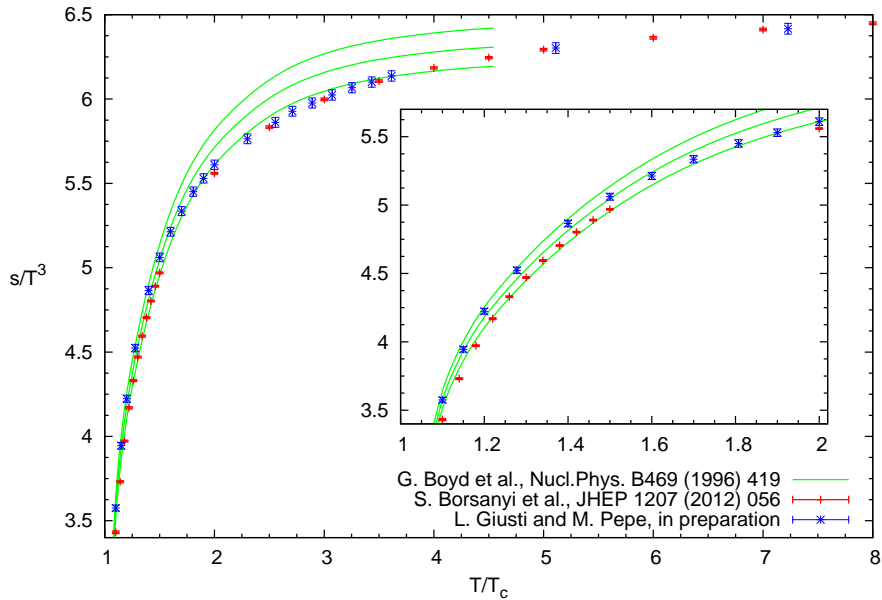}
\caption{
The temperature dependence of the dimensionless ratio $s/T^3$. The results obtained using 
eq.~(\protect\ref{EoS}) are compared with data available in the literature.\label{EoSplot}}
\end{figure}

In Fig.~\ref{EoSplot} we compare our preliminary data with the results available in the
literature. In the region 1-2.5 $T_c$ we find a discrepancy with \cite{Borsanyi:2012ve} and our data are
compatible with \cite{Boyd:1996bx}. At larger temperatures up to $T\simeq 7\; T_c$, our results agree
with those presented in \cite{Borsanyi:2012ve}. Work is in progress to clarify the above mentioned
discrepancy and to reach temperatures about 250 $T_c$.

\section{Conclusions and outlook}
We have presented results for the non-perturbative renormalization of the energy-momentum
tensor on the lattice. In particular, the renormalization factors of the traceless diagonal and
off-diagonal components of $T_{\mu\nu}$ have been calculated. An equation to compute the
renormalization factor of the trace is also presented and work is in progress to perform
that measurement. The physical relevance of the non-perturbative renormalization of the
energy-momentum tensor is discussed presenting the results of Monte Carlo simulations to
compute the Equation of State of the $SU(3)$ Yang-Mills theory.
The framework of shifted boundary conditions turns out to be very effective to investigate
the Yang-Mills theory at finite temperature. Work is in progress to include also dynamical
fermions. \\
{\it Acknowledgements.} M.P. wishes to thank K.~Szabo for the invitation to the Conference.

\end{document}